\documentclass[12pt]{iopart}

\usepackage{iopams}
\usepackage{setstack}
\usepackage{graphicx}
\usepackage{bm}
\usepackage{verbatim}
\DeclareMathAlphabet{\mathpzc}{OT1}{pzc}{m}{it}
\makeatletter
\DeclareRobustCommand{\text}{%
  \ifmmode\expandafter\text@\else\expandafter\mbox\fi}
\let\nfss@text\text
\def\text@#1{{\mathchoice
  {\textdef@\displaystyle\f@size{#1}}%
  {\textdef@\textstyle\f@size{#1}}%
  {\textdef@\textstyle\sf@size{#1}}%
  {\textdef@\textstyle \ssf@size{#1}}%
  \check@mathfonts
  }%
}
\def\textdef@#1#2#3{\hbox{{%
                    \everymath{#1}%
                    \let\f@size#2\selectfont
                    #3}}}
\makeatother

\begin{document}

%\title[Influence of the gravitational field produced by a global monopole in f(R) gravity on massive scalar fields]{Influence of the gravitational field produced by a global monopole in f(R) gravity on massive scalar fields}

\title[Abelian cosmic string in the Starobinsky model of gravity]{Abelian cosmic string in the Starobinsky model of gravity}

\author{J. P. Morais Gra\c ca$^{1}$}

\address{$^{1}$ Departamento de F\'{i}sica, Universidade Federal da Para\'{i}ba, Caixa Postal 5008, CEP 58051-970, Jo\~{a}o Pessoa, PB, Brazil}

\ead{jpmorais@gmail.com}

\begin{abstract}
In this paper, I analyze numerically the behaviour of the solutions corresponding to an Abelian string in the framework of the Starobinsky model. The role played by the quadratic term in the Lagrangian density $f(R) = R + \eta R^2$ of this model is emphasized and the results are compared with the corresponding ones obtained in the framework of Einstein's theory of gravity. I have found that the angular deficit generated by the string is lowered as the $\eta$ parameter increases, allowing a well-behaved spacetime for a large range of values of the symmetry-breaking scale.
\end{abstract}


%\msc{81Q05, 83C45, 83C57, 83C75}

%\preprint{AIP/123-QED}

\maketitle

%\begin{quotation}
%...
%\end{quotation}

%
%%%%%%%%%%%%%%%%%%%%%%%%%%%%%%%%%%%%%%%%%%%%%%%%%%%%%%%%%%%%%%%%%%%%%%%%%%%%%%%%%%%%%%%%%%%%%% Introduction
%
\section{Introduction}
Einstein's theory of gravity in its final form was presented in 1915, and since then several alternatives to replace it have been proposed. Among the arguments for such endevour, its non-renormalizability and the necessity of $ad$ $hoc$ ingredients to deal with the recent observation of an accelerated universe are of main importance. In the late 70's it was shown that the addition of curvature invariant squared terms in the Lagrangian density can improve the renormalizability of the theory \cite{Stelle:1976gc}, and recently the so-called $f(R)$ theories of gravity have deserved some attention due to the fact that it can model the observed acceleration of the universe without an $ad$ $hoc$ cosmological constant \cite{Sotiriou:2008rp}. 

Cylindrically symmetric solutions of $f(R)$ theories of gravity have been studied in \cite{Azadi:2008qu}\cite{Sharif:2012sv}\cite{Momeni:2009tk}, but due to the strong non-linearity of the theory solutions have been obtained only for very specific $f(R)$ models and mainly for asymptotic values. The most used procedure is to search for specific ideal metrics and then let the equations of motion give us the functional $f(R)$ form that allows for the existence of the proposed metrics. In an alternative Brans-Dicke approach this would mean to let the scalar potential $V(\phi)$ to start arbitrary and then let the equations of motion impose us which potential to use. This method is straightfoward and allows us to obtain some exact solutions for the theory, but the price to be paid is to deal with extremely complicated $f(R)$ functions (or unusual potentials, in a Brans-Dicke scenario). Here I will use a more direct approach and choose the $f(R)$ functional form from the beginning. The Starobinsky model of gravity is probably the most simple $f(R)$ theory of gravity (but far from trivial), where the Lagrangian density of the Einstein-Hilbert action is replaced by $f(R) = R + \eta R^2 $. In this model we obtain an accelerated stage in the early universe which is compatible with the recent results from Planck satellite \cite{Planck:2013jfk} and in this way it can be considered as a candidate to describe gravity at high energies. In spite of the simple polynomial form for this $f(R)$ theory, its field equations are highly non-linear and we can't find exact solutions, in general. 

Topological defects are stable solutions of classical fields where their stability is provided by topological arguments related to the vacuum manifold. They are expected to be formed in the early stages of the universe. They cannot be considered as primary sources of CMB anisotropy \cite{Ade:2013xla}, but in a inflationary scenario it contributes to the observed power spectrum. Cosmic strings seen also to play some role in which concerns the BB polarization obtained by BICEP2 experiment \cite{Ade:2014xna}. Thus it is natural that, if we believe that some effective theory will replace Einstein's gravity in this scenario, we should study the formation and behaviour of topological defects, in particular cosmic strings, in the context of these effective theories, such as $f(R)$ theories of gravity and others. A cosmic string is an example of a cylindrically symmetric topological defect. For a review on the subject, see \cite{Hindmarsh:1994re}\cite{Sakellariadou:2009ev}.  

In this paper I will study the Abelian Higgs model of cosmic strings in the Starobinsky model of gravity. Due to its squared Ricci scalar, we can claim that this correction term should play a fundamental role in the strong curvature regime, such as near compact objects or in the primordial universe era, where the Abelian Higgs model is expected to generate cosmic strings during some phase transition. This result is independent of the gravitational theory behind. This string model was first introduced by Nielsen and Olesen in 1973 as a field-theoretical model to mimic the features of the Nambu-Goto string, in that time a strong candidate to explain Hadronic physics \cite{Nielsen:1973cs}. Today quantum chromodynamics is the most used theory to study hadronic physics, but strings still play fundamental roles, as fundamental superstrings or fundamental strings. 

The gravitational effects due to the Abelian Higgs string model were first analytically studied in \cite{Vilenkin:1981zs} for an idealized cosmic string as a fixed background, and numerically in \cite{Garfinkle:1985hr}. In the later, the Einstein-Maxwell-Higgs equations were solved together, as it should be. Such cosmic strings are known to generate an asymptotic conical geometry and so they can, in principle, be observed by their astrophysical effects, such as gravitational lensing. As we will see, the magnitude of the conical geometry is directly related to the $\eta$ parameter of the gravitational correction term and the scale of symmetry breaking, and a parametrization on such parameters can be used to enlarge the possible ranges for the occurrence of the symmetry breaking. A complete classification of the string-like solutions in Einstein's gravity can be found in \cite{Christensen:1999wb}. Our main goal in this paper is to extend these previous results taking into account the $R^2$ term in the action, and to analyze how the string properties are affected by it.

This paper in organized in the following manner. In section 2 I will present our model and the field equations we must solve. In section 3 I will solve these equations numerically and present our results comparing it with our knowledge about the same model in Einstein's gravity. Finally in section 4 I will present our conclusions.

\section{The Model}

The action for a gravitating Abelian Higgs system in the Starobinsky model is given by

\begin{equation}
S = \int{d^4x \sqrt{|g|}(\frac{1}{2}D_\mu\Phi^* D^\mu \Phi - \frac{\lambda}{4}(\Phi^*\Phi - \nu^2)^2 - \frac{1}{4}F_{\mu\nu}F^{\mu\nu} + \frac{1}{16 \pi G}f(R))}
\label{action}
\end{equation}
where $f(R) = R + \eta R^2$, which means that the action of general relativity is corrected by a quadratic term in the Ricci scalar, $F_{\mu\nu}$ is the Abelian field strength, $\Phi$ is a complex scalar field with vacuum expectation value $\nu$ and $D_\mu = \nabla_\mu - i e A_\mu$ is the gauge covariant derivative. I am using units were $ c = \hbar = 1$. Note that if the parameter $\eta$ is null the above action reduces to the Abelian Higgs model in Einstein's gravity.

Because of the cylindrical symmetry of the source and due to the symmetry under boosts along the string axis, I will adopt a line element

\begin{equation}
ds^2 = - N^2(r) dt^2 + dr^2 + L^2(r) d\phi^2 + N^2(r) dz^2 ,
\end{equation}
and the usual Nielsen-Olesen ansatz for the unity flux string \cite{Nielsen:1973cs}

\begin{eqnarray}
\Phi(r) = \nu f(r) e^{i \phi} 
\\
A_{\mu} dx^\mu = \frac{1}{e} [1 - P(r)] d\phi .
\end{eqnarray}

This ansatz is chosen in such a way that the matter fields reach their vacuum expectation values at infinity when we impose the boundary conditions \cite{Nielsen:1973cs}. Varying the above action with respect to the scalar and gauge fields, we get the following set of field equations

\begin{eqnarray}
\frac{(N^2Lf')'}{N^2L} + (\lambda \nu^2 (1-f^2) - \frac{P^2}{L^2})f = 0
\\
\frac{L}{N^2}(\frac{N^2}{L}P')' - e^2 \nu^2 f^2 P = 0,
\end{eqnarray}
where $'$ means derivative with respect to the radial coordinate. The energy-momentum tensor for the matter fields is as usually given by $T^{\mu\nu} = \frac{2}{\sqrt{-g}}\frac{\delta S_{matter}}{\delta g_{\mu\nu}}$, and its components are

\begin{eqnarray}
& T_t^{\phantom{t}t} = - \epsilon_s - \epsilon_v - \epsilon_w - u \\
\nonumber
& T_r^{\phantom{r}r} = + \epsilon_s + \epsilon_v - \epsilon_w - u \\
\nonumber
& T_\phi^{\phantom{\phi}\phi} = - \epsilon_s + \epsilon_v + \epsilon_w - u \\
\nonumber
& T_z^{\phantom{z}z} = T_t^{\phantom{t}t},
\end{eqnarray}
where 

\begin{eqnarray}
& \epsilon_s = \frac{\nu^2}{2}f'^2,  \hspace{10pt}
\epsilon_v = \frac{P'^2}{2 e^2 L^2}, \hspace{10pt}
\epsilon_w = \frac{\nu^2 P^2 f^2}{2L^2} \hspace{10pt}and\hspace{10pt}
u = \frac{\lambda \nu^4}{4}(1-f^2)^2 .
\end{eqnarray}
The gravitational field equations are given by 

\begin{equation}
G_{\mu\nu}(1 + 2 \eta R) + \frac{1}{2} \eta g_{\mu\nu} R^2 - 2 \eta (\nabla_\mu \nabla_\nu - g_{\mu\nu} \Box)R = \kappa^2 T_{\mu\nu}
\end{equation}
where $G_{\mu\nu} = R_{\mu\nu} - \frac{1}{2} g_{\mu\nu} R$ and $\kappa^2 = 8 \pi G$, which is a set of fourth-order differential equations, but to avoid to deal with fourth-order differential equations we will treat the Ricci scalar as an independent field. This is an usual trick used in these kind of problems. The trace of the above equations is given by

\begin{equation}
-R + 6 \eta \Box R = \kappa^2 T .
\label{eq10}
\end{equation}
Note that eq. (\ref{eq10}) tell us that the Ricci scalar obeys a differential equation and not a purely algebraic one, as in Einstein's gravity. Before we deal with the components of the above differential equations, we should do some variable and field redefinitions to let all parameters dimensionless. To achieve this goal we will express all lengths in terms of the scalar characteristic length scale given by $1/\sqrt{\lambda \nu^2}$, so we will change our radial coordinate to a dimensionless coordinate $x = \sqrt{\lambda \nu^2} r$, together with the field redefinitions $L(x) = \sqrt{\lambda \nu^2} L(r)$ and $R(x) = R(r) / \lambda \nu^2$. We also introduce three new parameters, $\alpha = e^2 / \lambda $, $\gamma = 8 \pi G^2 \nu^2 $ and $ \xi = \eta \lambda \nu^2 $. Thus, we can write the equations of motion as

\begin{eqnarray}
&f'' = (\frac{P^2}{L^2} + f^2 - 1)f - \frac{(N^2L)'f'}{N^2L} ,
\label{eqqF}
\\
&P'' = \alpha f^2 P - \frac{L}{N^2}(\frac{N^2}{L})'P' ,
\label{eqqP}
\end{eqnarray}
for the fields and

\begin{eqnarray}
\label{eqqN}
N'' =  -\frac{1}{24} \frac{1}{\alpha N L^2 (1 + 2 \xi R)}[6 \xi \alpha L (4 R L N'^2+ R^2 L N^2- 4 R' L' N^2) 
\\
\nonumber
- \gamma N^2 (10 \alpha P^2 f^2 + \alpha (1 - f^2)^2 L^2 - 2 \alpha f'^2 L^2 + 6 P'^2)  \\
\nonumber
+ 4 \alpha L^2 (3 N'^2 + R N^2)],
\end{eqnarray}

\begin{eqnarray}
\label{eqqR}
R'' = \frac{1}{24} \frac{1}{N^2 L^2 \alpha \xi}[6 \xi \alpha L R(8 N' N L' + 4 N'^2 L + R L N^2) \\
\nonumber
-\gamma N^2(10 \alpha f'^2 L^2 - 2 \alpha P^2 f^2  + \alpha L^2(1 - f^2)^2+ 6 P'^2 ) \\
\nonumber
+ 4 \alpha(6 N' L N L' + 3 N'^2 L^2 + R N^2 L^2)],
\end{eqnarray}

\begin{eqnarray}
\label{eqqL}
L'' = \frac{1}{24} \frac{1}{L \alpha N^2(1 + 2 \xi R)}[6 \xi \alpha L(4 N'^2 L R - R^2 N^2 L - 8 N' N L' R \\
\nonumber
+ 8 R' N' L N - 4 R' L' N^2) -  \gamma N^2(\alpha (2 f'^2 L^2 + 14 P^2 f^2 - L^2 (1- f^2)^2) \\   
\nonumber
+ 18 P'^2) + 4 \alpha L(3 N'^2 L - R N^2 L - 6 N' N L')]
\end{eqnarray}
where the primes in the above equations means derivatives with respect to x. Equations (\ref{eqqF}) and (\ref{eqqP}) give us the dynamics of the matter fields, while eqs (\ref{eqqN}) to (\ref{eqqL}) give us the behaviour of the metric fields as well as of the Ricci scalar field. We must also impose the boundary conditions on the fields. As mentioned, in order to get string-like solutions the matter fields should reach their vacuum expectation values asymptotically, so we must impose the following boundary conditions

\begin{eqnarray}
f(0) = 0, \hspace{10pt} f(\infty) = 1 \\
\nonumber
P(0) = 1, \hspace{10pt} P(\infty) = 0,
\end{eqnarray}
where infinity should be understood as a limit and the boundary conditions on the origin are necessary for the regularity of the fields. The boundary conditions for the metric functions are

\begin{eqnarray}
L(0) = 0, \hspace{10pt} L'(0) = 1 \\
\nonumber
N(0) = 1, \hspace{10pt} N'(0) = 0 \\
\nonumber
R'(0) = 0, \hspace{10pt} R(\infty) = 0 
\end{eqnarray}
where the first four conditions are necessary for the regularity of the metric. The last two boundary conditions are trickier, since the Ricci scalar is not an independent field. The last one is actually imposed by the field equations, and $R'(0)=0$ is a good choice to set the smoothness of the Ricci scalar. 

\begin{figure}
\centering
\includegraphics[scale=1.0]{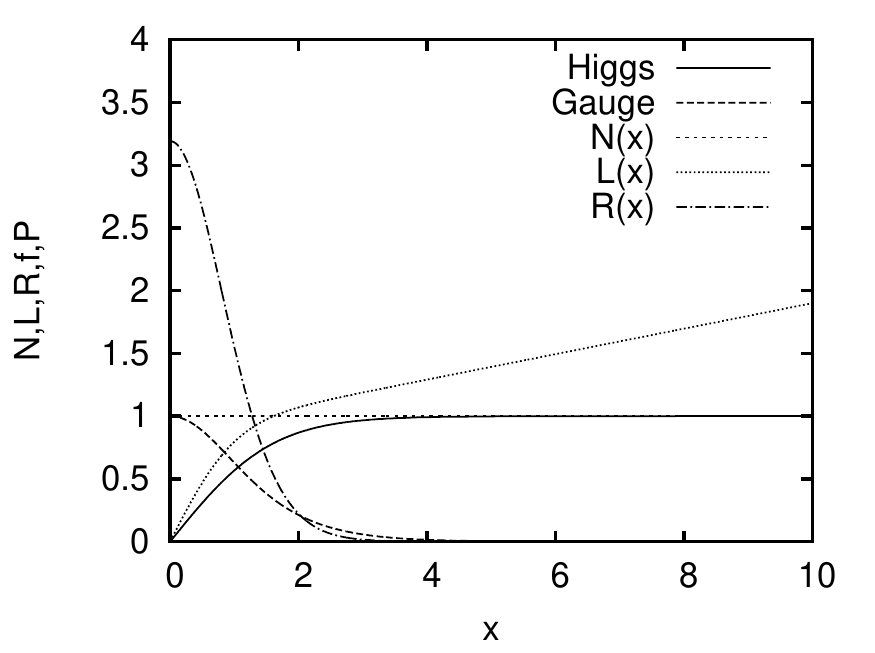}
\caption{A numerical solution for $\alpha = 2.0, \gamma = 1.8$ and $\xi = 0.001$ that resembles the Abelian string in Einstein's gravity. The inclination of the L component of the metric is directly related to the angular deficit due to the string.}
\label{fig:Einsteinstring}
\end{figure}

We now have to solve a three parameter set of five differential equations with boundary conditions. As we can see, these equations are highly non-linear and must be solved numerically. To better understand the string behaviour we should compare our results with the string behaviour in Einstein's gravity and how it is modified by the inclusion of the $\xi$ (related to the old $\eta$) parameter. Figure (\ref{fig:Einsteinstring}) shows a typical numerical example of a string in Starobinsky model with a low value for $\xi$. The behaviour for the metric and matter fields are in agreement with the well known results in Einstein's gravity (see \cite{Christensen:1999wb}), as it should be.

\subsection{The conical geometry}

The geometry generated by the cosmic string is an asymptotically flat conical geometry. Asymptotically, the metric fields present the following behaviour

\begin{eqnarray}
N(x \rightarrow \infty) = a,
\\
L(x \rightarrow \infty) = bx + c, \hspace{10pt} b \geq 0,
\end{eqnarray} 
where $a$, $b$ and $c$ are constants depending on $\alpha, \gamma$ and $\xi$. In the absence of gravity we expect $a=1, b=1$ and $c=0$. If we fix $\alpha$ and $\xi$, as we vary $\gamma$ we are in some way increasing the coupling between the matter fields and the gravitational field, and we expect the geometry be more and more affected, which is what really happens. Actually, the asymptotic intrinsic curvature is null, but the topology of the space is not trivial and its observational effects can be measured. This string-like solution possesses a planar deficit angle that can be expressed as

\begin{equation}
\Delta = 2\pi(1- L'(\infty)).
\end{equation}

It is worth calling attention to the fact that we can increase $\gamma$ until $\Delta$ reaches $2\pi$ (or $b = 0$). This value is called critical gamma, $\gamma_{cr}$, and for $\gamma > \gamma_{cr}$ the spacetime is not globally well-defined (for $b < 0$, there will be a maximum value for $L(x) > 0$).\footnote{In some dynamical scenarios, as in the topological inflation \cite{Vilenkin:1994pv}, it is possible to have a well-defined spacetime with $\gamma > \gamma_{cr}$. In this paper we will not worry about this specific issue, since we are using $\gamma_{cr}$ only as a parameter to quantify the geometry of the spacetime.}. Let's remember that $\gamma$ is directly related to the scale of the symmetry breaking, and so $\gamma_{cr}$ impose us a constraint on the maximum scale for the occurrence of it. In a real world scenario, we should expect that only a small value for the angular deficit will be allowed, otherwise its effects should already be noted (at least in present days), but to deal with critical values is useful for better understanding the properties of the theory.

In Einstein's gravity ($\xi = 0$) we can find the critical value for $\gamma$ as a function of $\alpha$, e. g. \cite{Brihaye:2000qr}

\begin{equation}
\gamma_{cr}(1.0) \approx	1.66, \hspace{10pt} \gamma_{cr}(3.0) \approx 2.2, \hspace{10pt} \text{and so on}.
\end{equation}

Our aim in the next section is to study how these results change as we increase $\xi$, or in other words, at what extent the correction due to the Starobinsky model becomes important. We will study how the matter and metric fields change and how the conical geometry is affected. 

\section{Numerical Results}

I have used a finite difference Newton-Raphson algorithm with adaptive grid scheme \cite{Archer1}\cite{Archer2} to construct the solutions numerically. Our estimated errors range from $10^{-9}$ to $10^{-12}$, sometimes even better. The limit $\xi \rightarrow 0$ corresponds to standard Einstein's gravity, but we can't use $\xi = 0$ in our analysis because the equations are not well defined at this value. The reason is that in Einstein's gravity the Ricci scalar obeys an algebraic relation with the energy-momentum tensor, not a differential equation. Nothing forbids us, however, to start with a small value for the parameter $\xi$ that mimics Einstein's gravity and increase it slowly. A typical string solution for $\xi = 0.001$ is plotted in figure (\ref{fig:Einsteinstring}).

\begin{figure}[htb]
\centering
\begin{tabular}{@{}cc@{}}
\includegraphics[scale=0.8]{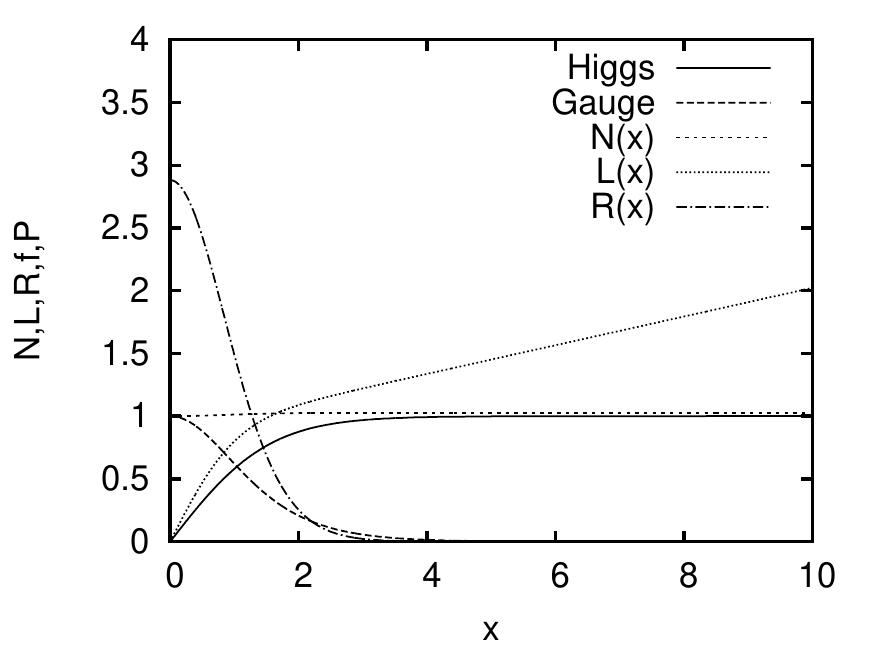} &
\includegraphics[scale=0.8]{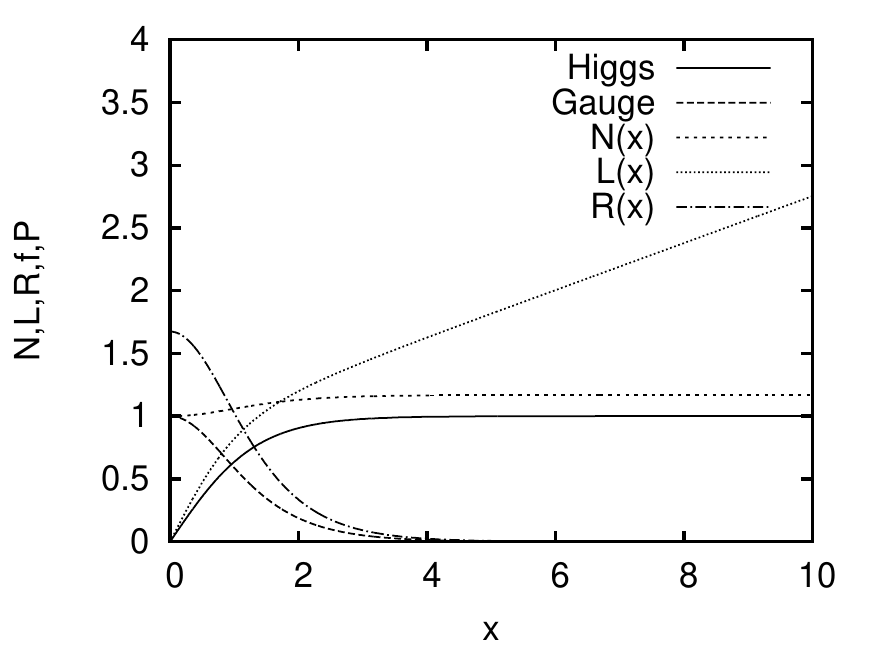} \\
\includegraphics[scale=0.8]{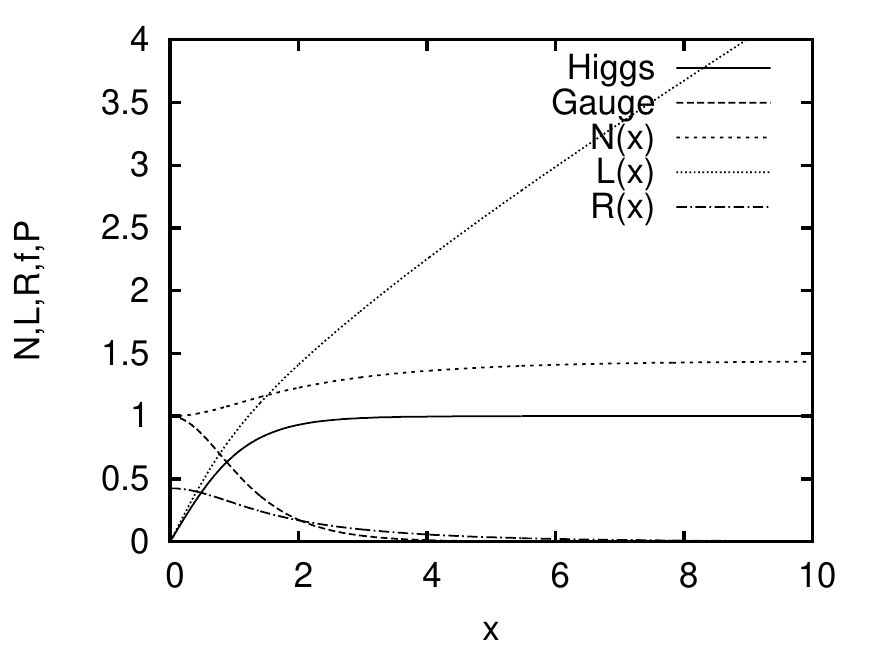} &
\includegraphics[scale=0.8]{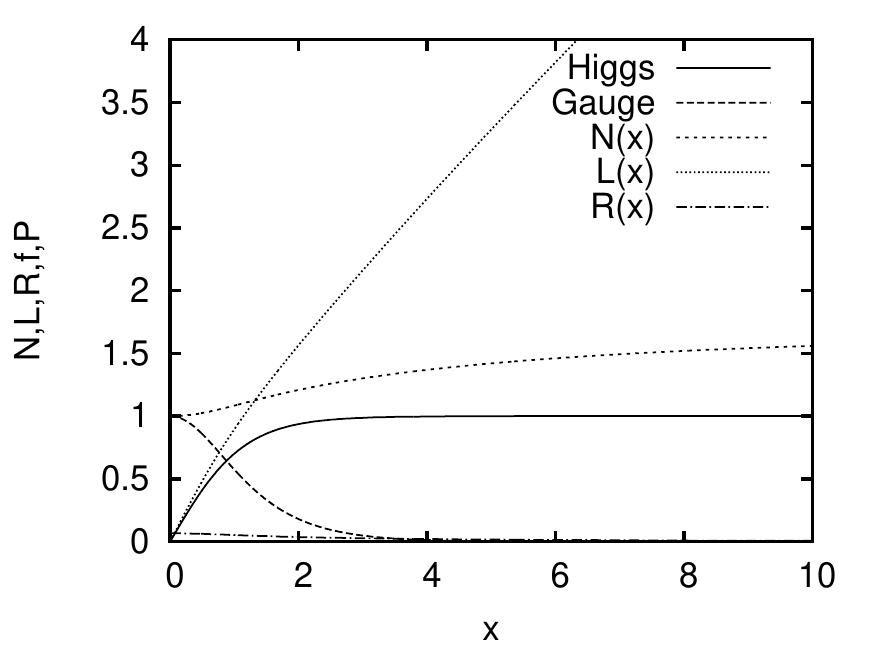}
\end{tabular}
\caption{Numerical solutions for $\alpha = 2.0, \gamma = 1.8$. On the top left and right, respectivelly, $\xi = 0.01$ and $\xi = 0.1$, and on the bottom left and right, respectivelly, $\xi = 1.0$ and $\xi = 10.0$. The gravitational effects are weaker for lager $\xi$ values.}
\label{fig:howstringchanges}
\end{figure}

In this section we will study how the behaviour of the metric and matter fields are influenced by the addition of the $R^2$ term in the action, or how the Abelian string in the Starobinsky model of gravity differs from the one in standard Einstein's gravity. We will proceed changing the parameter $\xi$ keeping the other two parameters, $\gamma$ and $\alpha$, fixed. Taking $\gamma = 1.8$ and $\alpha = 2.0$ fixed, for example, we can plot the metric behaviour due to the string in the range $\xi = 0.01$ to $\xi = 10$. The plots are in figure (\ref{fig:howstringchanges}) starting upper left with $\xi = 0.01$. We can note that, as we increase $\xi$, the matter fields do not change considerably from their usual value, but the Ricci scalar field decreases to almost zero. The changes in the gravitational fields are better noticed by the change in the inclination of the $L(x)$ component of the metric. As I exposed in the last section, this inclination is directly related to the angular deficit generated by the string and so, as we increase the $\xi$ parameter, the geometry becomes less and less conical.

\begin{figure}
\centering
\includegraphics[scale=1.0]{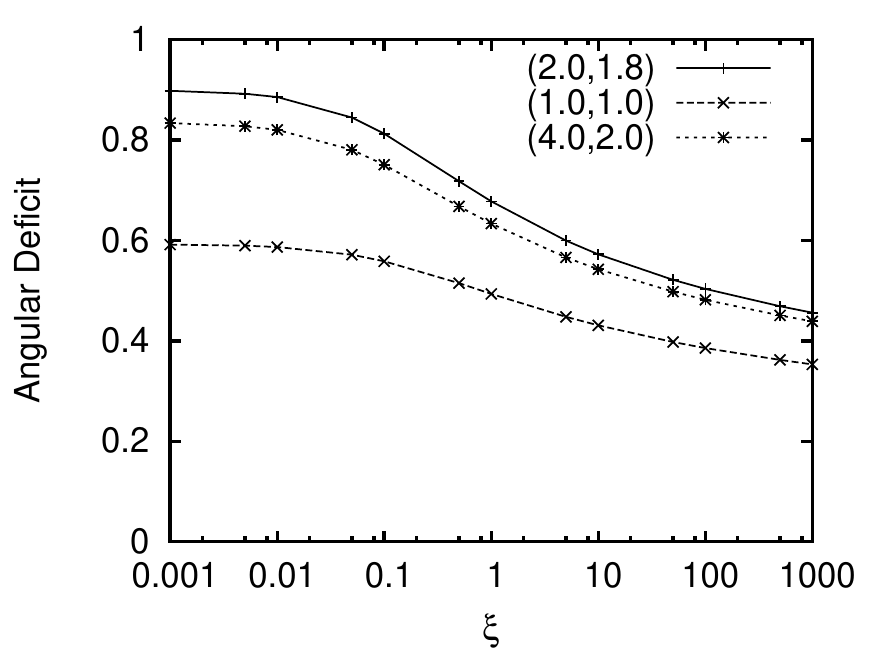}
\caption{How the angular deficit changes in 3 differents $(\alpha, \gamma)$ parameter-sets for $\xi$ ranging from $0.001$ to $1000$. The vertical axis measures the angular deficit over $2\pi$}
\label{fig:Angulardeficit}
\end{figure}

By itself, the angular deficit is an important quantity due to its consequences on observational aspects of cosmic strings. It is responsible for effects such as gravitational lensing and also affects the anisotropy of the cosmic microwave background, just to cite two examples \cite{Hindmarsh:1994re}. I have shown that the angular deficit changes as we vary $\xi$ for the set $(\alpha = 2.0, \gamma = 1.8)$ in figure (\ref{fig:howstringchanges}), but we must analyze its behaviour also for other parameter sets. In figure (\ref{fig:Angulardeficit}) I plot how the angular deficit changes in 3 differents $(\alpha, \gamma)$ parameter sets, namely $(2.0, 1.8)$, $(1.0, 1.0)$ and $(4.0, 2.0)$ with $\xi$ ranging from $0.001$ to $1000$, where unity on the vertical axis means an angular deficit of $2\pi$ and the zero value means no angular deficit at all. We can clearly see that as $\xi$ grows, the angular deficit shrinks for all parameter sets.

\begin{figure}
\centering
\includegraphics[scale=1.0]{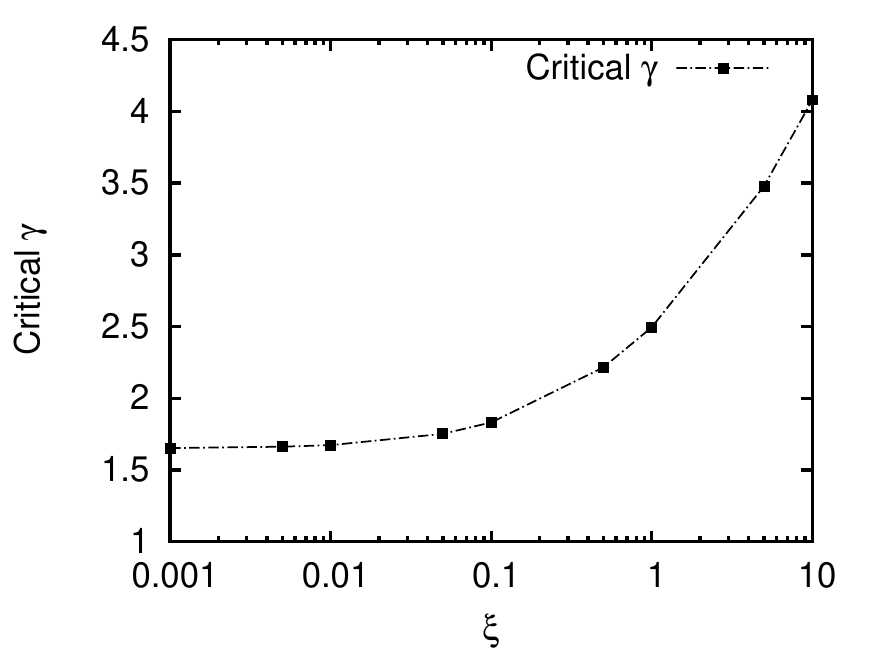}
\caption{Critical value for $\gamma(\alpha = 1.0)$ as function of $\xi$. The inclusion of an $R^2$ term turns the geometry less conical.}
\label{fig:critical}
\end{figure}

This change on the angular deficit is important, among the reasons above mentioned, because it can avoid the parameter $\gamma$ to reach its critical value. In a cosmological scenario, if we believe that cosmic strings exists, we should expect the strings to put a boundary on the the scale of the symmetry breaking, or vice versa, otherwise they can give rise to space-times not globally well defined. The presence of an $R^2$ correction term in the Einstein-Hilbert action can avoid this kind of problem, allowing larger values for $\gamma_{cr}$. 

This is what the above examples show us. As the $R^2$ term becomes more relevant, the theory becomes less conical, and we should expect the maximum value for the symmetry breaking scale to be bigger. As an example, the critical value for $\gamma(\alpha = 1.0)$ in Einstein's gravity is about 1.66, but as we increase $\xi$ this value starts to grow. In figure (\ref{fig:critical}) I plot a graph for $\gamma_{cr}(\alpha = 1.0, \xi)$ which show us that for large values of $\xi$, the new scale where the symmetry breaking scale reaches $\gamma_{cr}$ is much higher than the original one. This can allows large values for the scale of symmetry breaking before we reach observational constraints.

\begin{figure}[htb]
\centering
\begin{tabular}{@{}cc@{}}
\includegraphics[scale=0.8]{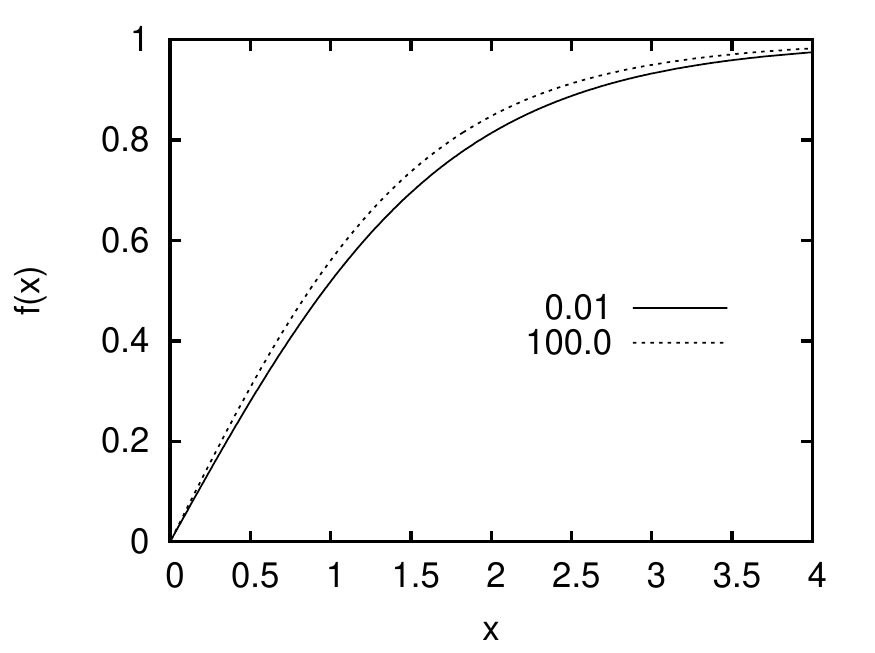} &
\includegraphics[scale=0.8]{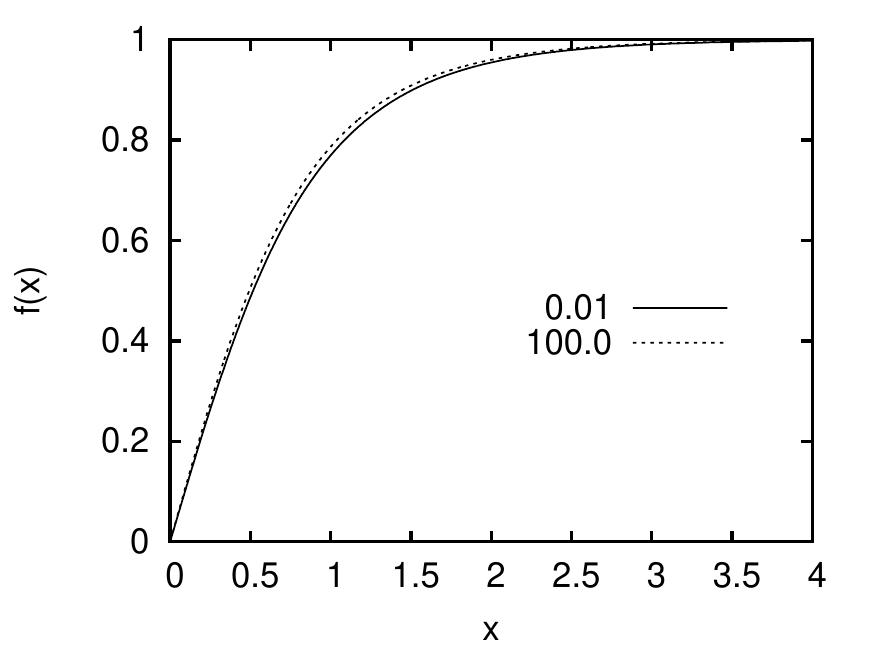} \\
\includegraphics[scale=0.8]{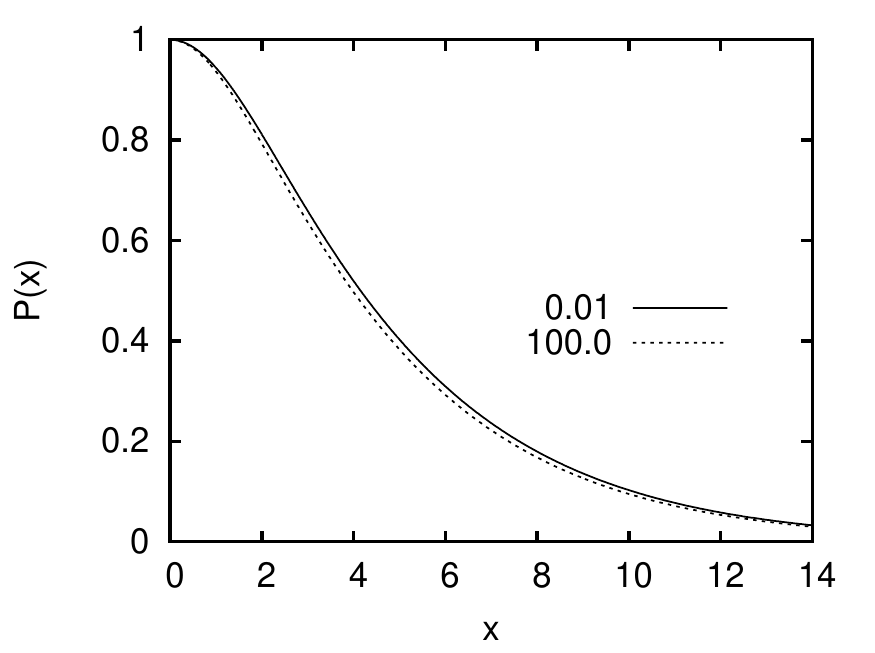} &
\includegraphics[scale=0.8]{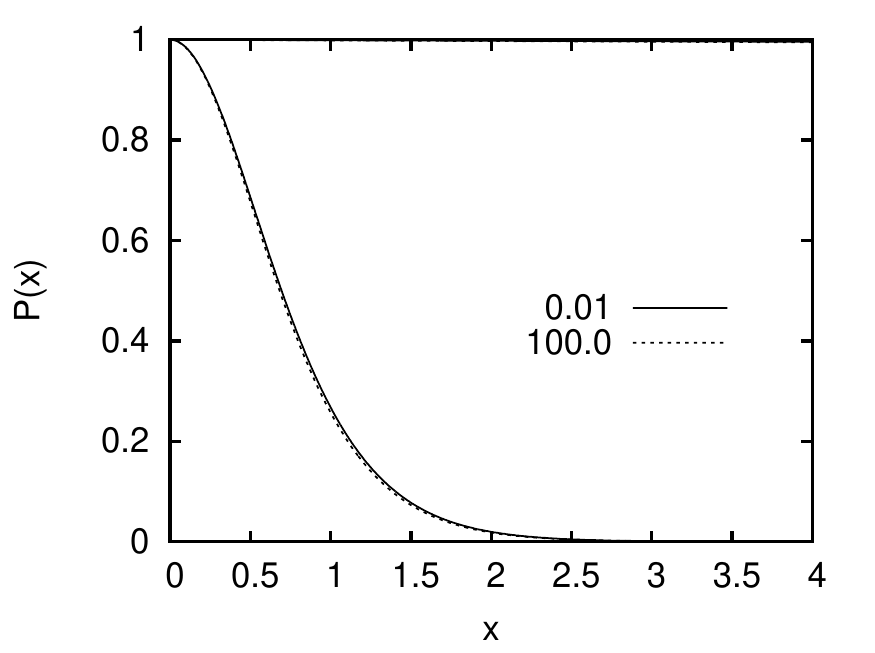} \\
\end{tabular}
\caption{This plots show how the Gauge ($P(r)$) and Higgs ($f(R)$) fields change as we vary $\xi$. We can see that the matter fields are not significantly affected. The left column was calculated with $\alpha = 0.1$, $\gamma = 0.5$ and the right column with $\alpha = 10$, $\gamma = 0.5$.}
\label{fig:matterfields}
\end{figure}

\begin{table}[]
\centering
\caption{Values for $M_{in}$, as given by Eq. (\ref{eqnEnergy}), and angular deficit. The first pair of values are for ($\alpha = 1$, $\gamma = 1$) and the second pair are for ($\alpha = 4$, $\gamma = 2$).}
\label{tableEnergy}
\begin{tabular}{p{0.15\textwidth}p{0.15\textwidth}p{0.15\textwidth}p{0.15\textwidth}p{0.15\textwidth}p{0.15\textwidth}}
\hline
 $\xi$  & Energy & $\delta \phi / 2\pi$  & Energy & $\delta \phi / 2\pi$ \\ \hline
 0.001  & 1.1711 & 0.5918  & 0.8527	& 0.8339 \\
 0.01	& 1.1781 & 0.5872  & 0.8641 & 0.8206 \\
 0.1	& 1.2063 & 0.5588  & 0.9061 & 0.7508 \\
 1.0	& 1.2307 & 0.4938  & 0.9315 & 0.6335 \\
 10  	& 1.2262 & 0.4310  & 0.9251 & 0.5428 \\
 100	& 1.2153 & 0.3858  & 0.9136 & 0.4820 \\
 1000	& 1.2067 & 0.3532  & 0.9061 & 0.4390 
\end{tabular}
\end{table}

We will now study how the matter fields are affected by the $R^2$ term. In figure (\ref{fig:matterfields}) I plot the Higgs and gauge fields for different values of the $\xi$ parameter. As we can see, the matter fields themselves aren't significantly influenced by the addition of an $R^2$ term, and so I will not go further on this aspect of the theory. In any case, the string-like structure is preserved.

The way the energy of the string differs from its standard value was also calculated. I will follow \cite{Brihaye:2000qr} and define the energy (mass per unit length) as

\begin{eqnarray}
G \mathcal{M} = \frac{\gamma}{8} M_{in} = \frac{\gamma}{8} \int^{\infty}_{0} dx NL \left( (f')^2 + \frac{(P')^2}{\alpha L^2} + \frac{P^2 f^2}{L^2} + \frac{1}{2}(1-f^2)^2 \right).
\label{eqnEnergy}
\end{eqnarray}

The calculated values for $M_{in}$ are listed in table (\ref{tableEnergy}) for 2 different $(\alpha, \gamma)$ parameter sets. In the BPS limit, where the Higgs mass is equal the gauge mass, $M_{in}$ is equal one in a flat spacetime.

\subsection{A brief discussion on vacuum}

To study the Starobinsky model in vacuum I assume $\gamma = 0$, and so the metric equations decouple from the matter fields. Cylindrically symmetric and static spacetimes in $f(R)$ theories have been analytically studied in \cite{Azadi:2008qu}, where the authors pointed out the possibility of the occurrence of angular deficit even in vacuum for $R = 0$. This deficit angle would be related with integration constants that depends on the $f(R)$ functional form. I performed numerical analysis for $\gamma = 0$ and $\xi$ from $0$ to $1000$, and found no evidence of the existence of an angular deficit beyond the numerical error limits. A preliminary numeric analysis on more general polynomial functions for $f(R)$, such as $f(R) = R + \alpha R^2 + \beta R^3 + \eta R^4$ also shows no evidence for a deficit angle in the flat asymptotic regime. 

\section{Conclusions and Perspectives}

In this paper I have studied how the Abelian Higgs model for a cosmic string is affected when we go from Einstein's gravity to the Starobinsky model of gravity. This gravity model can be seen as a first correction to Einstein's gravity in some effective theory, and so it is important that we study classical solutions also on these effective theories, specially when these classical solutions play fundamental roles on strong gravity regimes, as in the primordial universe era. 

The Abelian Higgs model in Einstein's gravity can be parametrized by 2 parameters ($\gamma$,$\alpha)$, where the first is related to the symmetry breaking scale and the second one is related with the internal structure of the string. In the Starobisky model of gravity, where an $\eta R^2$ term is added to the Einstein-Hilbert action, the same model gets a new parameter, $\xi$, related to the old $\eta$ parameter and so to the value of the correction. We now have a 3 parameter set and a more general theory to work with.

The conical geometry generated by such strings can constrains the maximum value for the symmetry breaking scale because for $\gamma > \gamma_{cr}$ the whole spacetime is no longer regular in some models. We have shown that the introduction of a new parameter reduces this constraint and allows larger values for this scale. In a scenario where this constraint appears to be violated in Einstein's gravity, we can propose to replace Einstein's gravity by a more general theory, such as the Starobinsky model of gravity. For this particular model, I have explicitly calculated the new $\gamma_{cr}$ for some parameter sets. I have also calculated the behaviour of the metric fields and compared to similar behaviour in Einstein's gravity. I also found no evidence of an angular deficit in the limit when $\gamma = 0$, where the gravitational equations decouple from matter, even for large values for the parameter $\xi$. 

I have also shown how the internal properties of the Abelian Higgs model are changed when we go to the Starobisnky model of gravity. I have shown that the string-like structure is preserved, and also that the scalar and gauge fields are not significantly affected by such a replacement. I have also calculated how the string energy (or mass per area) changes as the correction term gets stronger. 

It would be interesting to compare our model with the Abelian Higgs model on other $f(R)$ theories of gravity, specially one with more terms on a polynomial expansion. We are currently following this path. Also, models for non-Abelian cosmic strings are highly interesting and we expect to compute how our results differs in a non-Abelian model. As the calculations has shown that the interior of the comic string is barely affected by the $R^2$ term, I believe that the corrections will follow the same pattern obtained here.

\ack This work was supported by CAPES Fellowship. I would like to acknowledge Betti Hartmann and Valdir B. Bezerra for valuable discussions concerning this paper. I also would like to acknowledge the kind hospitality at Jacobs University Bremen where this work was partially done.

%
%%%%%%%%%%%%%%%%%%%%%%%%%%%%%%%%%%%%%%%%%%%%%%%%%%%%%%%%%%%%%%%%%%%%%%%%%%%%%%%%%%%%%%%%%%%%%% thebibliography
%

%
%%%%%%%%%%%%%%%%%%%%%%%%%%%%%%%%%%%%%%%%%%%%%%%%%%%%%%%%%%%%%%%%%%%%%%%%%%%%%%%%%%%%%%%%%%%%%%
%

\section*{References}
\begin{thebibliography}{99}
\bibitem{Stelle:1976gc} 
  K.~S.~Stelle,
  %``Renormalization of Higher Derivative Quantum Gravity,''
  Phys.\ Rev.\ D {\bf 16}, 953 (1977).
\bibitem{Sotiriou:2008rp} 
  T.~P.~Sotiriou and V.~Faraoni,
  %``f(R) Theories Of Gravity,''
  Rev.\ Mod.\ Phys.\  {\bf 82}, 451 (2010)
  [arXiv:0805.1726 [gr-qc]].
\bibitem{Azadi:2008qu} 
  A.~Azadi, D.~Momeni and M.~Nouri-Zonoz,
  %``Cylindrical solutions in metric f(R) gravity,''
  Phys.\ Lett.\ B {\bf 670}, 210 (2008)
  [arXiv:0810.4673 [gr-qc]].
\bibitem{Sharif:2012sv} 
  M.~Sharif and S.~Arif,
  %``Static cylindrically symmetric interior solutions in f(R) gravity,''
  Mod.\ Phys.\ Lett.\ A {\bf 27}, 1250138 (2012)
  [arXiv:1302.1191 [gr-qc]].
\bibitem{Momeni:2009tk} 
  D.~Momeni and H.~Gholizade,
  %``A note on constant curvature solutions in cylindrically symmetric metric $f(R)$ Gravity,''
  Int.\ J.\ Mod.\ Phys.\ D {\bf 18}, 1719 (2009)
  [arXiv:0903.0067 [gr-qc]].	
\bibitem{Planck:2013jfk} 
  P.~A.~R.~Ade {\it et al.} [Planck Collaboration],
  %``Planck 2013 results. XXII. Constraints on inflation,''
  Astron.\ Astrophys.\  {\bf 571}, A22 (2014)
  [arXiv:1303.5082 [astro-ph.CO]].
\bibitem{Ade:2013xla} 
  P.~A.~R.~Ade {\it et al.} [Planck Collaboration],
  %``Planck 2013 results. XXV. Searches for cosmic strings and other topological defects,''
  Astron.\ Astrophys.\  {\bf 571}, A25 (2014)
  [arXiv:1303.5085 [astro-ph.CO]].
\bibitem{Ade:2014xna} 
  P.~A.~R.~Ade {\it et al.} [BICEP2 Collaboration],
  %``Detection of $B$-Mode Polarization at Degree Angular Scales by BICEP2,''
  Phys.\ Rev.\ Lett.\  {\bf 112}, no. 24, 241101 (2014)
  [arXiv:1403.3985 [astro-ph.CO]].		
\bibitem{Hindmarsh:1994re} 
  M.~B.~Hindmarsh and T.~W.~B.~Kibble,
  %``Cosmic strings,''
  Rept.\ Prog.\ Phys.\  {\bf 58}, 477 (1995)
  [hep-ph/9411342].
\bibitem{Sakellariadou:2009ev} 
  M.~Sakellariadou,
  %``Cosmic Strings and Cosmic Superstrings,''
  Nucl.\ Phys.\ Proc.\ Suppl.\  {\bf 192-193}, 68 (2009)
  [arXiv:0902.0569 [hep-th]].	
\bibitem{Nielsen:1973cs} 
  H.~B.~Nielsen and P.~Olesen,
  %``Vortex Line Models for Dual Strings,''
  Nucl.\ Phys.\ B {\bf 61}, 45 (1973).
\bibitem{Vilenkin:1981zs} 
  A.~Vilenkin,
  %``Gravitational Field of Vacuum Domain Walls and Strings,''
  Phys.\ Rev.\ D {\bf 23}, 852 (1981).
\bibitem{Garfinkle:1985hr} 
  D.~Garfinkle,
  %``General Relativistic Strings,''
  Phys.\ Rev.\ D {\bf 32}, 1323 (1985).
\bibitem{Christensen:1999wb} 
  M.~Christensen, A.~L.~Larsen and Y.~Verbin,
  %``Complete classification of the string - like solutions of the gravitating Abelian Higgs model,''
  Phys.\ Rev.\ D {\bf 60}, 125012 (1999)
  [gr-qc/9904049].	
\bibitem{Vilenkin:1994pv} 
  A.~Vilenkin,
  %``Topological inflation,''
  Phys.\ Rev.\ Lett.\  {\bf 72}, 3137 (1994)
  [hep-th/9402085].  
\bibitem{Brihaye:2000qr} 
  Y.~Brihaye and M.~Lubo,
  %``Classical solutions of the gravitating Abelian Higgs model,''
  Phys.\ Rev.\ D {\bf 62}, 085004 (2000)
  [hep-th/0004043].	
\bibitem{Archer1} 
  U. Ascher, J. Christiansen and R. D. Russell,
	Math. Comput. {\bf 33} 659 (1979) 
\bibitem{Archer2} 
  U. Ascher, J. Christiansen and R. D. Russell,
	ACM Trans. Math. Softw {\bf 7} 209 (1981) 
	
\end{thebibliography}
\end{document}